\documentclass{ws-hepph08}

\newcommand{\de}{d}     
\newcommand{\ms}{\mskip 1.5mu}
\newcommand{\cdott}{{\mskip -1.5mu} \cdot {\mskip -1.5mu}}

\allowdisplaybreaks[4]

\begin{document}

\title{Semi-inclusive processes \\ at low and high transverse momentum}

\author{\underline{Alessandro Bacchetta}$^1$, Daniel Boer$^2$, 
Markus Diehl$^3$, and Piet J. Mulders$^2$}
 
\address{$^1$Jefferson Lab, 12000
Jefferson Ave, Newport News, VA 23606, USA
\\
$^2$Vrije Universiteit Amsterdam,
1081 HV Amsterdam, The Netherlands
\\
$^3$Deutsches Elektronen-Synchroton DESY,
22603 Hamburg, Germany}

\begin{abstract}
This talk reports on recent work where 
we studied the connection between the description of semi-inclusive DIS at
high transverse momentum (based on collinear factorization) and low transverse
momentum (based on transverse-momentum-dependent factorization). 
We used power counting to determine
the leading behavior of the structure functions 
at intermediate transverse momentum in the two descriptions. 
When the power behaviors are different, two distinct mechanisms are present and
there can be no
matching between them. When the power behavior is the same, the two descriptions
must match. An explicit calculation however shows that for some observables 
this is not the case, suggesting that the
transverse-momentum-dependent-factorization  
description beyond leading twist is 
incomplete. 
\end{abstract}

\bodymatter

\section{Introduction}
The cross section for polarized semi-inclusive DIS can be written in terms of 18
structure functions~\cite{Bacchetta:2006tn}. 
Each of them depends, among other variables, on 
the square of the transverse momentum of the
outgoing hadron, $P_{h\perp}^2$, with respect to the virtual photon
direction. For theoretical considerations, 
it is often preferable to consider the transverse momentum $q_T^2
\approx  P_{h\perp}^2/z^2$.
The problem involves three scales, namely
the scale of nonperturbative QCD dynamics, which we represent by the
nucleon mass $M$, the transverse momentum $q_T$, and the photon 
virtuality $Q$, which we
require to be large compared with $M$.

At high $q_T$ ($q_T \gg M$) the structure functions can be described using
collinear factorization, i.e., in terms
of collinear distribution and fragmentation functions together with
perturbative radiation.
At low-$q_T$ ($q_T \ll Q$) the structure functions can be described using
Collins--Soper TMD factorization~\cite{Collins:1981uk,Ji:2004wu}, i.e., 
in terms  
transverse-momentum-dependent (TMD)
parton distribution and fragmentation functions.
The low- and high-$q_T$ domains
overlap for $M \ll q_T \ll Q$ (intermediate $q_T$), 
where both descriptions can hence be
applied.

Studying the relation between the high-$q_T$ and low-$q_T$ regimes is
important both from the theoretical and phenomenological point of
view. We observe that in some cases the calculations in the two regimes have
to give the same result at intermediate $q_T$, i.e., they have to match. 
If this does not occur, we can make the important 
conclusion that there is some flaw in the formalism. We observe in other cases 
that the two calculations describe different mechanisms and therefore 
do not have to
match. Both of them have to be taken into consideration independently in the
overlap region.

\section{Matches and mismatches: general discussion}

To assess whether the high-$q_T$ and low-$q_T$ calculations have to
match or not in the intermediate-$q_T$ region it is sufficient to study the
power behavior of the structure functions in the two regimes. 

It is important to
realize that the power expansions are done in two different ways in the two
descriptions. 
At low $q_T$, first we expand in $(q_T/Q)^{n-2}$ and neglect terms
with $n$ bigger than a certain value (so far, analyses have been carried out
only up to $n=3$, i.e., twist-3). To study the behavior
at intermediate $q_T$ we further expand in $(M/q_T)^k$. Conversely, at high
$q_T$ we first expand in $(M/q_T)^n$ (also in this case, analyses are available
up to $n=3$, i.e., twist-3). 
To study the intermediate-$q_T$ region, we further expand 
in $(q_T/Q)^{k-2}$.


We can encounter two different situations. For simplicity, we will refer to
them as type-I and type-II observables.

\vspace{-0.25cm}
\subsection{Type-I observables}

Consider, e.g., a structure functions described by two contributions
\begin{equation}
F={A} \, \frac{M^2}{M^2 + q_T^2} + {B}\,
\frac{q_T^2} {Q^2} \frac{M^2}{M^2 + q_T^2}. 
\end{equation} 
At low transverse momentum, term $B$ is neglected from the very beginning
because it is of order  $(q_T/Q)^{2}$ (twist-4). The remaining term is 
\begin{equation}
F^{\text{twist-2}}_{\rm low} = {A} \, \frac{M^2}{M^2 + q_T^2} = 
A\,\frac{M^2}{q_T^2} + {\cal O}\left(\frac{M^4}{q_T^4}\right), 
\end{equation} 
where the second step identifies the leading term at $q_T \gg M$.

At high transverse momentum, both terms $A$ and $B$ are twist-2 and are taken
into consideration. However, the second term is neglected if a further
expansion in $q_T/Q$ is 
performed, to study the regime of intermediate transverse momentum, i.e.,
\begin{equation}
F^{\text{twist-2}}_{\rm high} = A\,\frac{M^2}{q_T^2}+B\,\frac{M^2}{Q^2}
%
= A\,\frac{M^2}{q_T^2} + {\cal O}\left(\frac{q_T^2}{Q^2}\right).
\end{equation} 
Therefore, the {\em leading} terms in the two expansions are the same.
In this case, the
calculations at high and low transverse momentum  
must yield exactly the same result at intermediate transverse
momentum~\cite{Collins:1984kg,Ji:2006ub}.  
If a
mismatch occurs, it means that one of the calculations is incorrect or
incomplete.

\vspace{-0.25cm}
\subsection{Type-II observables: expected mismatches}

Consider the example of a structure functions composed by two terms
\begin{equation}
F={A} \, \frac{M^4}{M^4 + q_T^4} + {B}\,
\frac{q_T^2} {Q^2} \frac{M^2}{M^2 + q_T^2}. 
\end{equation} 
At low transverse momentum, term $B$ is neglected from the very beginning
because it is of order  $(q_T/Q)^{2}$ (twist-4). What is left is
\begin{equation}
F^{\text{twist-2}}_{\rm low} = {A} \, \frac{M^4}{M^4 + q_T^4} = 
A\,\frac{M^4}{q_T^4} + {\cal O}\left(\frac{M^8}{q_T^8}\right), 
\end{equation} 
where in the second step we expanded in $M/q_T$.

At high transverse momentum, the term $A$ is now twist-4 and it is usually
neglected. Only the second term is kept and gives
\begin{equation}
F^{\text{twist-2}}_{\rm high} = B\,\frac{q_T^2}{Q^2}\,\frac{M^2}{q_T^2}
%
\end{equation} 
In this case, if the calculations at high and low transverse momentum are 
performed at their 
respective leading twist, they 
correspond to two different contributions to the cross section and will 
not lead to the same result at intermediate transverse momentum. In order
 to
``match'', the calculations would have to be carried out in both regimes up to the
sub-subleading order. We could
call this situation an ``expected mismatch'', since it is simply due to the
difference  between the two expansions.

\section{Matches and mismatches: semi-inclusive DIS case}

In Tab.~\ref{tab:overview} we list the power behavior of
the structure functions at intermediate transverse momentum, 
as obtained from the limits of the low-$q_T$ and high-$q_T$ calculation. For
details of the calculation, we refer to Ref.~\refcite{Bacchetta:2008xw}. In the
names of the structure functions, the
first and
second subscript 
respectively specifies the polarization of the beam
and the target.  
When present, the third subscript refers to
the polarization of the photon.

\begin{table}
\tbl{\label{tab:overview} 
  Behavior of SIDIS structure functions in
  the region $M\ll q_T \ll Q$.  Empty fields indicate that
  no calculation is available (in this case, twist $4$
  indicates observables that are zero when calculated up to
  twist-three accuracy).  Yes/no in parentheses: expected answers based on
  analogy, rather than actual calculation.
}
{\renewcommand{\arraystretch}{1.3}
\begin{tabular}{|l|cc|cc|c|c|} \hline
 & \multicolumn{2}{|c|}{low-$q_T$ calculation}
 & \multicolumn{2}{|c|}{high-$q_T$ calculation} & power & exact \\
observable & twist & power & twist & power
 & match & match\\ \hline
$F^{}_{UU,T}$ & 2  & $1/q_T^{2}$
              & 2  & $1/q_T^{2}$ & yes & yes \\
$F^{}_{UU,L}$ & 4 &
              & 2 & $1/Q^{2}$ &  & \\
$F^{\cos\phi_h}_{UU}$ & 3 & $1/(Q\ms q_T)$
                      & 2 & $1/(Q\ms q_T)$ & yes & no \\
$F^{\cos 2\phi_h}_{UU}$ & 2 & $1/q_T^{4}$
                        & 2 & $1/Q^{2}$ & no & \\
$F^{\sin\phi_h}_{LU}$ & 3 & $1/(Q\ms q_T)$
                      & 2 & $1/(Q\ms q_T)$ & yes & (no) \\
$F^{\sin\phi_h}_{UL}$ & 3 & $1/(Q\ms q_T)$
                      & & & (yes) & (no) \\
$F^{\sin 2\phi_h}_{UL}$ & 2 & $1/q_T^{4}$ & & & (no) &  \\
$F^{}_{LL}$ & 2 & $1/q_T^{2}$
            & 2 & $1/q_T^{2}$ & yes & yes \\
$F^{\cos\phi_h}_{LL}$ & 3 & $1/(Q\ms q_T)$
                      & 2 & $1/(Q\ms q_T)$ & yes & no \\
$F^{\sin(\phi_h-\phi_S)}_{UT,T}$ & 2 & $1/q_T^{3}$
                                 & 3 & $1/q_T^{3}$ & yes & yes \\
$F^{\sin(\phi_h-\phi_S)}_{UT,L}$ & 4 & 
                      & 3 & $1/(Q^2\ms q_T)$ &  & \\
$F^{\sin(\phi_h+\phi_S)}_{UT}$ & 2 & $1/q_T^{3}$
                               & 3 & $1/q_T^{3}$ & yes & (yes) \\
$F^{\sin(3\phi_h-\phi_S)}_{UT}$
                      & 2 & $1/q_T^{3}$
                      & 3 & $1/(Q^{2}\ms q_T)$ & no & \\
$F^{\sin\phi_S}_{UT}$ & 3 & $1/(Q\ms q_T^{2})$
                      & 3 & $1/(Q\ms q_T^{2})$ & yes & (no) \\
$F^{\sin(2\phi_h-\phi_S)}_{UT}$
                      & 3 & $1/(Q\ms q_T^2)$
                      & 3 & $1/(Q\ms q_T^2)$ & yes & (no)\\
$F^{\cos(\phi_h-\phi_S)}_{LT}$ & 2 & $1/q_T^{3}$ & & & (yes) & (yes) \\
$F^{\cos\phi_S}_{LT}$ & 3 & $1/(Q\ms q_T^{2})$ & & & (yes) & (no) \\
$F^{\cos(2\phi_h-\phi_S)}_{LT}$
                      & 3 & $1/(Q\ms q_T^{2})$ & & & (yes) & (no) \\ 
\hline
\end{tabular}}
\end{table}

In summary, the calculation at high $q_T$ is done using standard collinear
factorization, as done in, e.g., Ref.~\refcite{Mendez:1978zx,Koike:2006fn} 
and in Ref.~\refcite{Eguchi:2006mc}  
for the subleading-twist
sector. 
To obtain the power behavior at intermediate $q_T$, we need to perform
an expansion in $q_T/Q$. The calculation has no fundamental difficulties and
allows us to fill in the third column of Tab.~\ref{tab:overview}.
The blank entries correspond to the structure functions that have not yet been
computed in the high-transverse-momentum regime. 

\begin{figure}
\center
\includegraphics[width=9cm]{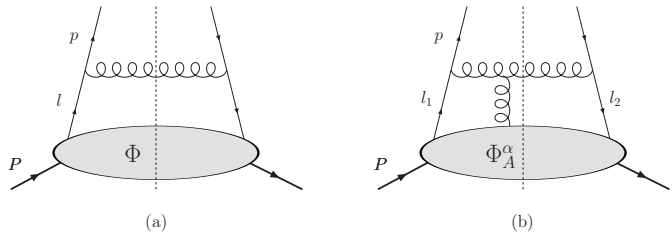}
\caption{\label{f:rungdistA} Example diagrams for the calculation of
  the high-$p_T$ behavior TMD parton distribution functions. $\Phi_A^{\alpha}$
  represent the quark-gluon-quark correlator. The dashed lines represent the
  final-state cut.} 
\end{figure}

The calculation at low $q_T$ is done using
TMD factorization~\cite{Collins:1981uk,Ji:2004wu}. The behavior of
the TMD functions at intermediate transverse momentum can be calculated
perturbatively by considering diagrams as the ones depicted in
Fig.~\ref{f:rungdistA}. 
We calculated the power behavior of all twist-2 and twist-3 TMD functions, 
which allowed us to
fill in the second column of Tab.~\ref{tab:overview}. Two structure functions
cannot be calculated as they require twist-4 contributions, which are beyond
the current limits of the TMD factorization framework. 

The fourth column of Tab.~\ref{tab:overview} is obtained by 
comparing the second and third column.
The structure functions with a ``yes'' are type-I observables, those with a
``no'' are type-II. The values in parentheses are expectations based on
analogy with similar structure functions, 
since the high-$q_T$ calculations are not available.

Beside studying the power behavior, 
we also calculated the explicit form of some of the TMD functions at
intermediate transverse momentum, namely the
ones requiring only the evaluation of diagrams analogous to that of
Fig.~\ref{f:rungdistA}(a). 
A calculation of the Sivers function, requiring the
evaluation of diagrams like that of Fig.~\ref{f:rungdistA}(b),   
was already performed in
Refs.~\refcite{Ji:2006ub,Koike:2007dg}. 

The explicit calculations allows us to check if for type-I observables 
the
explicit expressions obtained from high and low transverse momentum exactly
match or not. The results are listed in the fifth column of
Tab.~\ref{tab:overview}. The entries in parentheses are conjectures based on
analogy rather than actual calculation.

\vspace{-0.25cm}
\subsection{Type-I structure functions}

For type-I structure functions (``yes'' in the column ``power match''), 
we know from power counting that the two calculations 
describe the same physics and should therefore exactly match.
In these cases, 
the high-$q_T$ calculation corresponds to the perturbative tail of the
low-$q_T$ effect. The two mechanisms need not be distinguished. Using
resummation it should be possible to construct expressions for these
observables that are valid at any $q_T$, as was done for the Drell-Yan analog
of $F_{UU,T}$ in  Ref.~\refcite{Collins:1984kg}. 

Only five of these structure
functions have been calculated explicitly: $F_{UU,T}$, $F_{LL}$ and
$F^{\sin(\phi_h-\phi_S)}_{UT,T}$ (Sivers structure function) 
present an exact matching~\cite{Ji:2006ub,Koike:2007dg}, while in
our work we showed that $F^{\cos\phi_h}_{UU}$ and $F^{\cos\phi_h}_{LL}$ do not
match.  In analogy
to these results, we expect that also $F^{\sin(\phi_h+\phi_S)}_{UT}$ (Collins
structure function) and
$F^{\cos(\phi_h-\phi_S)}_{LT}$ will match exactly, while problems will occur
with all the others, since they are twist-3 in the low-$q_T$ regime, and the
TMD factorization formalism is probably complete only at twist 2.

The structure function $F^{\sin(\phi_h-\phi_S)}_{UT,T}$, related to the Sivers 
function, is an example of a match between high- and low-$q_T$. Some of the
consequences of the calculation are:
\begin{itemize}
\item{the leading (twist-3, in this case) 
      contribution of the high-$q_T$ calculation corresponds to the tail of the
      Sivers function at intermediate $q_T$, it is not a competing effect and
      should not be summed to the Sivers function; }
\item{it is conceivable to construct an expression that extends the 
    high-$q_T$ calculation to $q_T \approx M$, through a smooth
    merging into the Sivers function;}
\item{since the structure function falls as $1/q_T^3$, 
     it is safe to use $q_T$-weighted asymmetries to extract the Sivers
     function.}
\end{itemize}

As an example of a mismatch  we consider the structure function
$F_{UU}^{\cos\phi_h}$, 
related to the Cahn effect. We show in this case the main steps of the
calculation to explain the nature of the problem.
In the low-$q_T$ formalism, the expression for this observable
is~\cite{Bacchetta:2006tn} 
\begin{align} 
F_{UU}^{\cos\phi_h}
= \frac{2M}{Q}\,\mathcal{C}\biggl[
   &- \frac{\hat{\bm{h}}\cdott \bm{k}_T^{}}{M_h}
   \biggl(x  h\, H_{1}^{\perp } 
   + \frac{M_h}{M}\,f_1 \frac{\tilde{D}^{\perp }}{z}\biggr)
\nonumber
\\ &
   - \frac{\hat{\bm{h}}\cdott \bm{p}_T}{M}
     \biggl(x f^{\perp } D_1
   + \frac{M_h}{M}\,h_{1}^{\perp } \frac{\tilde{H}}{z}\biggr)\biggr],
\end{align} 
where the convolution means
\begin{align} 
\mathcal{C}\bigl[ w f D \bigr] = 
\sum_a x\ms e_a^2 
& \int \de^2 \bm{p}_T\,  \de^2 \bm{k}_T^{}\, \de^2 \bm{l}_T^{}\,
\delta^{(2)}\bigl(\bm{p}_T - \bm{k}_T^{} + \bm{l}_T^{} + \bm{q}_T \bigr)
\nonumber
\\
& \times w(\bm{p}_T,\bm{k}_T^{})\,
f^a(x,p_T^2)\, D^a(z,k_T^2)\, U(l_T^2) \,.
\label{e:conv}
\end{align} 
The term $U$ denotes the so-called soft factor. It is obtained in the
factorization proof for twist-two observables. Here we assume we can 
use it also for twist-three observables.
The terms with
$h_{1}^{\perp}$ and $H_{1}^{\perp}$ fall off as $1/p_T^3$ or $1/k_T^3$ and
are power suppressed compared to the terms with
$f^\perp$ and $\tilde{D}^\perp$ when $q_T \gg M$.  For intermediate
$q_T$ we therefore have
\begin{equation}
  \label{FUUcos-expand}
F_{UU}^{\cos\phi_h} =
-\frac{2 q_T}{Q} \sum_a x\ms e_a^2\,
\biggl[
  x f^{\perp a} (x,q_T^2)\, \frac{D_1^a(z)}{z^2}
- f_1^a(x)\, \frac{\tilde{D}^{\perp a}(z,q_T^2)}{z}
\biggr]
\end{equation}
at leading power.  In this case there
is no leading contribution from the soft factor taken at large
transverse momentum. The tail of the functions at  $q_T\gg M$ can be
calculated perturbatively and yields 
\begin{align} 
x f^{\perp q}(x,p_T^2)
&= \frac{\alpha_s}{2 \pi^2}\,
   \frac{1}{2 \bm{p}_T^2}\,
   \biggl[\ms \frac{L(\eta^{-1})}{2}\, f_1^q(x)
     + \bigl(P'_{qq} \otimes f_1^q + P'_{qg} \otimes f_1^g \bigr)(x)
   \biggr] \,,
\nonumber
\\
\frac{\tilde{D}^{\perp q}(z,k_T^2)}{z} &=
- \frac{\alpha_s}{2 \pi^2}\,
  \frac{1}{2 z^2\ms {\bm k}{}_T^2}\,
  \biggl[\ms \frac{L( \eta_{\smash{h}}^{-1} )}{2}\, D_1^q(z)
   - 2 C_F D_1^q(z)
\nonumber
\\& \hspace{3cm}
   + \bigl(D_1^q \otimes P'_{qq} + D_1^g \otimes P'_{gq}\bigr)(z)
  \biggr] \,,
\end{align}
where
$L(y) =
2 C_F \ln y - 3 C_F \,,$
and $P_{qq}'$, $P_{gq}'$, $P_{qg}'$ are kernels 
specific to the functions under consideration. 
The parameters $\eta$ and $\eta_h$ are related to the choice of a nonlightlike
gauge (Wilson line) 
in the calculation of the functions and fulfill the relation $\sqrt{\eta
  \eta_h } = q_T^2/Q^2$. Putting these ingredients together 
we arrive at
\begin{align} 
\label{e:high_FUUcosphi}
F_{UU}^{\cos\phi_h}
&= - \frac{1}{Q\ms q_T}\, \frac{\alpha_s}{2\pi^2 z^2}
  \sum_a x\ms e_a^2\,
  \biggl[f_1^a(x)\,D_1^a(z)\, L\biggl( \frac{Q^2}{q_T^2} \biggr)
\nonumber
\\ & \qquad 
+ f_1^a(x)\, \bigl( D_1^a \otimes P_{qq}'
                    +  D_1^a \otimes P_{gq}'\bigr)(z)
\nonumber
\\ & \qquad
+ \bigl( P_{qq}' \otimes f_1^a
       + P_{qg}' \otimes f_1^g\bigr)(x)\, D_1^a(z)
- 2 C_F\ms f_1^a(x) \,D_1^a(z)
\biggr] \,.
\end{align} 
This expression differs from the one obtained at high $q_T$ 
by the last term $2 C_F f_1^a(x)
\,D_1^a(z)$.  
At this point we are forced to conclude that the description of twist-3
structure functions is incomplete in the 
TMD-factorization formalism. However, it is interesting to note that adding
a term $f_1^a(x)\,D_1^a(z)z^{-2} U(q_T^2)/2$  within brackets in
Eq.~\eqref{FUUcos-expand} would be sufficient to cure this 
problem.   It is however not clear how such an expression 
        would be obtained from a factorized formula.

\vspace{-0.25cm}
\subsection{Type-II structure functions}

For type-II structure functions (``no'' in the column ``power match'')
the
low-$q_T$ and high-$q_T$ calculations at leading order
 pick up two different components of the
full structure function. They therefore describe two different mechanisms and
do not match. 

An example of a type-II observable is the structure function
$F^{\cos2\phi_h}_{UU}$, related at low $q_T$ to the Boer--Mulders
function~\cite{Boer:1997nt}. Some 
studies of this structure functions
have recently appeared~\cite{Barone:2008tn,Zhang:2008ez}. However, 
some considerations have to be kept
in mind:
\begin{itemize}
\item{the leading contribution from the high-$q_T$ calculation (often referred
    to as a pQCD or radiative correction) is a competing
    effect that has to be taken into account;}
\item{it is at present not possible to construct an expression that extends the 
    high-$q_T$ calculation to $q_T \approx M$, since this requires a smooth
    merging into unknown twist-4 contributions in TMD factorization;} 
\item{Using $q_T$-weighted asymmetries to extract the Boer--Mulders function
    is {\em not} a good idea, since the high-$q_T$ mechanism dominates the
    observable;}
\item{a solution to the above problems could be to consider observables
  that are least sensitive to the effect of radiative corrections, for
  instance by considering specific combinations of structure functions.}
\end{itemize}
We stress that the above considerations apply not only to semi-inclusive DIS,
but also to Drell--Yan and $e^+ e^-$
annihilation~\cite{Boer:2008fr}. Drell--Yan data have been already used to
extract~\cite{Zhang:2008nu} 
the Boer--Mulders function, without taking into account 
radiative corrections, while the
extraction~\cite{Anselmino:2007fs}  of the Collins function
from  $e^+ e^-$ relies on the cancellation
of radiative effects through the construction of suitable experimental
observables~\cite{Abe:2005zx,Boer:2008fr}.


%
%
%

%


\end{document}